\documentclass[letterpaper]{article} 
\usepackage{aaai25}  
\usepackage{times}  
\usepackage{helvet}  
\usepackage{courier}  
\usepackage[hyphens]{url}  
\usepackage{graphicx} 
\usepackage{natbib}  
\usepackage{caption} 
\usepackage{amsfonts}
\usepackage{graphicx}
\usepackage{multirow}
\usepackage{booktabs}
\usepackage[utf8]{inputenc}
\usepackage[linesnumbered,ruled,vlined]{algorithm2e}
\usepackage{amsmath}
\usepackage{amssymb}  
\usepackage{algpseudocode}

\usepackage{newfloat}
\usepackage{listings}
\usepackage{float}
\usepackage{placeins}
\usepackage{xcolor}

\DeclareCaptionStyle{ruled}{labelfont=normalfont,labelsep=colon,strut=off} 
\floatstyle{ruled}
\newfloat{listing}{tb}{lst}{}
\floatname{listing}{Listing}
\title{NexusIndex: Integrating Advanced Vector Indexing and Multi-Model Embeddings for Robust Fake News Detection}

\author {
    Solmaz Seyed Monir\textsuperscript{\rm 1}, 
    Dongfang Zhao\textsuperscript{\rm 2}
}
\affiliations {
    \textsuperscript{\rm 1}University of Washington, USA\\
    \textsuperscript{\rm 2}University of Washington, USA\\
    {solmazsm@uw.edu, dzhao@cs.washington.edu}
}

\begin{document}
\maketitle
\begin{abstract}
The proliferation of fake news on digital platforms has underscored the need for robust and scalable detection mechanisms. Traditional methods often fall short in handling large and diverse datasets due to limitations in scalability and accuracy. In this paper, we propose NexusIndex, a novel framework and model that enhances fake news detection by integrating advanced language models, an innovative FAISSNexusIndex layer, and attention mechanisms. Our approach leverages multi-model embeddings to capture rich contextual and semantic nuances, significantly improving text interpretation and classification accuracy. By transforming articles into high-dimensional embeddings and indexing them efficiently, NexusIndex facilitates rapid similarity searches across extensive collections of news articles. The FAISSNexusIndex layer further optimizes this process, enabling real-time detection and enhancing the system's scalability and performance. Our experimental results demonstrate that NexusIndex outperforms state-of-the-art methods in efficiency and accuracy across diverse datasets.
\end{abstract}

%
\section{Introduction}
\subsubsection{Motivation.} Recently, pre-trained language models (PLMs) such as BERT \cite{devlin2018bert}, RoBERTa \cite{liu2019roberta} have significantly advanced the field of Information Retrieval (IR) by capturing intricate token-wise correlations and producing high-quality representations through fine-tuning \cite{zhou2024fine}. These models have demonstrated significant improvements in the relevance and precision of retrieval results. However, detecting fake news requires not only effective retrieval but also a nuanced understanding of context and semantics to differentiate between true and false content. In the context of fake news detection, IR plays a crucial role by retrieving related content and cross-verifying sources, which helps the model assess the credibility of a given article. By retrieving relevant news articles, IR supports classification models in distinguishing between real and fake news based on similarities or discrepancies in the retrieved information. Traditional methods, often relying on basic linguistic features and shallow machine learning techniques, have shown limitations in effectively identifying fake news due to their inability to capture complex patterns and semantic subtleties \cite{zhang2024reinforced}.

\subsubsection{Challenges.} Despite the significant progress made with PLMs in the field of IR, there are still numerous challenges to address. Traditional cross-encoder methods \cite{zhang2022resnest}, which involve concatenating a query and document pair for detailed relevance assessment, are computationally demanding and unsuitable for large-scale applications due to the exponential growth of query-document pairs. In contrast, bi-encoder techniques, which independently embed queries and documents into single vectors \cite{gao2022long, wang2022simlm} within a shared dense semantic space, help mitigate efficiency issues by allowing offline document embeddings and streamlined online relevance computation \cite{zhou2024fine}. However, these bi-encoder methods face limitations due to the constrained capacity of single-vector representations, leading to performance drawbacks compared to more sophisticated cross-encoder models.

Previous works in fake news detection have leveraged both labeled and unlabeled data to enhance detection accuracy. Methods like Positive and Unlabeled Learning (PUL) and One-Class Learning (OCL) have shown improved performance by effectively utilizing both types of data \cite{bekker2020learning}. Additionally, hybrid weakly supervised learning techniques have been used to provide labels to unlabeled data, combined with deep learning models for detection \cite{rahimi2021bidirectional}. However, these approaches often rely on simpler text representation models such as Bag-of-Words or Doc2Vec and do not fully exploit the capabilities of advanced pre-trained language models (PLMs). Recent methods have shown that PLMs can significantly enhance fake news detection by capturing deeper contextual and semantic nuances. These models allow for better representation of the subtleties in language, which are often crucial for differentiating between real and fake news content. Approaches leveraging PLMs have demonstrated improved performance over traditional text representation techniques by understanding the fine-grained semantics in news articles \cite{radford2019language, devlin2018bert, liu2019roberta}.

A further challenge in fake news detection is illustrated by the WELFake dataset, where the most frequent keywords in both fake and real news articles are nearly identical. This overlap complicates the task of distinguishing between the two, as both types of articles often cover similar topics, particularly those related to high-profile political figures and events. This finding underscores the limitations of relying solely on keyword frequency for fake news detection. To address these challenges, we propose the \textbf{NexusIndex framework}. This methodology allows us to overcome the limitations of both traditional IR methods and simple keyword analysis. Unlike keyword frequency, embeddings capture the context in which words are used. For instance, keyword-based methods often rely on the frequency of certain words, such as ``Trump,'' ``election,'' or ``government,'' as seen in both fake and real news articles. While the keyword ``Trump'' might appear frequently in both fake and real news, its contextual usage could differ significantly, leading to misclassification when relying solely on keyword frequency. Embeddings allow the model to comprehend these differences, even if the keyword itself is the same. Similarity Search is used to efficiently retrieve similar articles based on their embeddings. This approach helps the model assess how closely a fake news article resembles a real news article in terms of content and context, rather than just shared keywords. Finally, we train a neural network on the multi-Model embeddings to classify articles as fake or real. This method leverages the deep semantic information captured by the embeddings, improving the model's ability to detect subtle differences between fake and real news.

\paragraph{Proposed Work}
Our approach leverages advanced multi-model embeddings to capture a broader range of contextual and semantic nuances, addressing the information bottleneck inherent in simpler models. The NexusIndex framework introduces \textbf{novel vector database} techniques for organizing and retrieving high-dimensional embeddings, significantly enhancing scalability and retrieval speed. These innovations techniques \textbf{NexusIndex} as an effective framework for real-time fake news detection, offering substantial improvements in both accuracy and efficiency. This leads to the following research questions:
\begin{enumerate}
\item How can multi-model embeddings improve the accuracy of fake news detection in large-scale information retrieval systems?

\item How does the integration of an optimized indexing system and attention mechanisms with multi-model embeddings enhance the accuracy and scalability of real-time fake news detection?
\end{enumerate}

\subsubsection{Contribution.} 
To address the complexities of fake news detection, NexusIndex pioneers innovative methods in embedding utilization and information retrieval. Our key contributions are: 
\begin{itemize}
\item We propose \textbf{NexusIndexModel II}, a novel architecture that integrates multi-model embeddings with an innovative \textbf{FAISSNexusIndex layer}, enabling rapid and precise similarity searches directly within the neural network.
\item We introduce a novel attention mechanism within \textbf{NexusIndexModel II} that enhances the model’s ability to focus on the most relevant features within the embeddings. This attention mechanism, combined with the \textbf{FAISSNexusIndex layer}, significantly improves the model’s scalability and accuracy in \textbf{indexing} and handling large-scale datasets for fake news detection.

\end{itemize}
\section{Related Work}
Our study introduces NexusIndexModel, a novel approach that enhances fake news detection through an Attention Layer and a FAISSNexusIndex for efficient similarity search. While \cite{kwon2013prominent} examines features that drive rumor spread in social media, and \cite{Whitehouse2022} integrates Wikidata knowledge into PLMs for improved detection, our method focuses on optimizing semantic embeddings and refining retrieval processes. Unlike prior approaches that combine external knowledge and user behavior features, NexusIndex significantly improves accuracy and robustness across diverse datasets. While prior studies, such as \cite{dun2021kan}, have advanced fake news detection by combining external knowledge and user behavior features, our NexusIndex model introduces a novel methodology. By optimizing semantic embeddings and refining the retrieval process through efficient similarity searches.

Fake news detection has been explored with various state-of-the-art deep learning. \cite{Shu2017} introduced a text-based framework, while \cite{pan2018content} used knowledge graphs to detect false information. However, these methods struggle with unstructured data. Our proposed \textbf{NexusIndexModel} overcomes these challenges by utilizing high-dimensional semantic embeddings and efficient vector database indexing for rapid and accurate similarity searches. Unlike models relying solely on explicit knowledge graphs, NexusIndex excels in \textbf{recognizing semantic patterns}, providing a more scalable solution for diverse datasets. Additionally, \cite{hu2024mpl} introduced the state-of-theart Multi-modal Prompt Learning (MPL) framework, which enhances early fake news detection by leveraging the multimodal pre-trained model CLIP. While MPL enhances multimodal analysis, NexusIndex surpasses it with efficient similarity searches, integrating semantic embeddings, and advanced retrieval. (Reimers and Gurevych 2019) developed SentenceBERT to enhance semantic embeddings for similarity tasks. Rather than relying solely on cosine similarity, our approach integrates Facebook AI Similarity Search (FAISS) \cite{Johnson2019, douze2024faiss} indexing for efficient high-dimensional searches, improving speed and accuracy in large-scale datasets.
\begin{figure*}[h]
\centering
\includegraphics[width=1\textwidth]{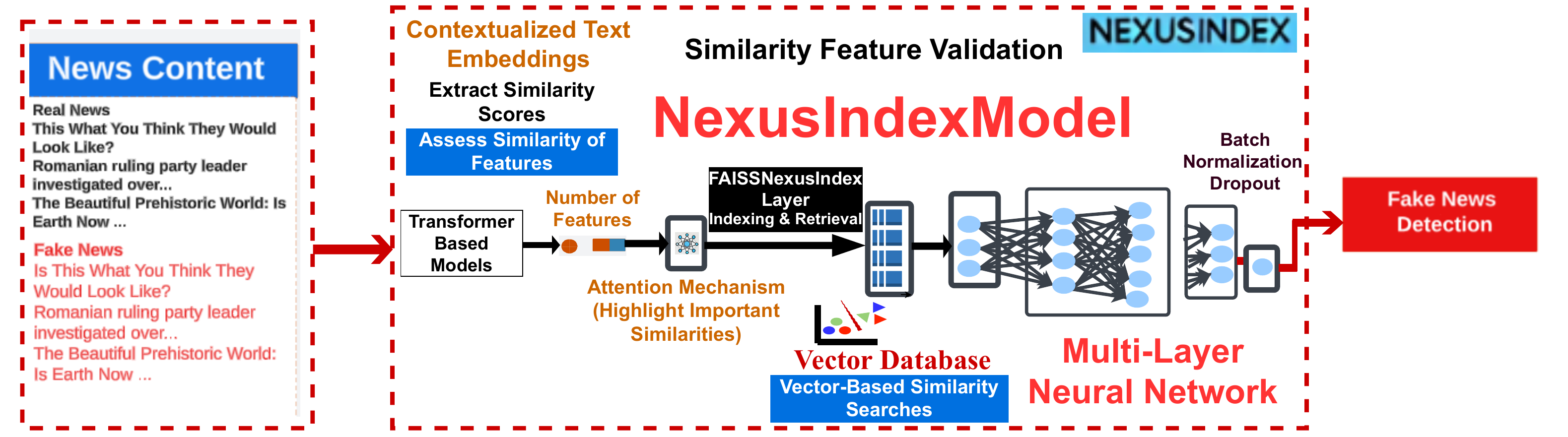}
\caption[width=0.8\textwidth]{Proposed a novel NexusIndex framework for fake news detection, which integrates Multi-Model to extract semantic embeddings. The framework features an innovative FAISSNexusIndex layer that efficiently indexes and retrieves relevant information. An attention mechanism is proposed to prioritize key similarities, which are then integrated for effective classification. The model is refined and evaluated through grid search to identify the best-performing configuration. Additionally, it leverages a vector database to ensure scalability and rapid similarity searches.
}
\label{fig:framework}
\end{figure*}
\section{Problem Statement}
Given a dataset \( D \) containing news articles labeled as either real (\( y = 1 \)) or fake (\( y = 0 \)), the objective is to accurately classify these articles using advanced multi-model embeddings and efficient similarity search techniques. For each article \( x_i \in D \), embeddings \( e_i^{\text{model}} \) are obtained using Multi-Model, where \( \text{model} \in \{ \text{BERT}, \text{RoBERTa}, \text{GPT}, \text{DistilBERT} \} \). These embeddings are indexed in a vector database \( I \) to enable rapid retrieval of similar articles based on the \( L_2 \) distance. A specialized layer, termed \textbf{FAISSNexusIndex Layer}, is integrated directly into the neural network to perform efficient similarity searches during both training and inference. The output of the \textbf{FAISSNexusIndex Layer} for an embedding \( e_i \) is given by the set of distances \( D_{i} = f_{\text{FAISS}}(e_i) = \text{top-k}\left(\|e_i - e_j\|_2\right) \) for all \( j \in I \). Additionally, an attention mechanism is utilized to refine the embeddings by prioritizing the most relevant information, with the refined embedding \( \tilde{e_i} \) computed as \( \tilde{e_i} = a(e_i) \odot e_i \). These refined embeddings are then used as input features in a neural network \( f_\theta \), defined as \( h_1 = \text{ReLU}(W_1 D_i + b_1) \), \( h_2 = \text{ReLU}(W_2 h_1 + b_2) \), and \( \hat{y} = \sigma(W_3 h_2 + b_3) \). The aim is to minimize the binary cross-entropy loss \( L(\theta) = -\frac{1}{N} \sum_{i=1}^{N} \left[ y_i \log(\hat{y}_i) + (1 - y_i) \log(1 - \hat{y}_i) \right] \) to optimize the network parameters \( \theta \), thereby enhancing the accuracy of fake news detection.
\section{Methodology}
To achieve robust and scalable fake news detection, we have designed the NexusIndex framework. The methodology involves several key steps.
\subsection{NexusInde Framework}
The NexusIndex framework is designed to enhance fake news detection by integrating advanced neural network layers that leverage both attention mechanisms and vector-based indexing for efficient similarity searches.
Given an input embedding $\mathbf{x}$, the FAISS layer retrieves the top-5 nearest neighbors' distances, denoted as $\mathbf{d} = \text{FAISS}(\mathbf{x})$. These distances are then processed through the NexusIndexModel as follows: the first hidden layer is computed as $\mathbf{h}_1 = \text{Dropout}(\text{ReLU}(\text{BatchNorm}(\mathbf{W}_1 \mathbf{d} + \mathbf{b}_1))$, with a dropout probability of $p=0.5$. The second hidden layer is obtained as $\mathbf{h}_2 = \text{Dropout}(\text{ReLU}(\text{BatchNorm}(\mathbf{W}_2 \mathbf{h}_1 + \mathbf{b}_2))$, again with a dropout probability of $p=0.5$. Finally, the output is calculated using the sigmoid activation function, $\hat{y} = \sigma(\mathbf{W}_3 \mathbf{h}_2 + \mathbf{b}_3)$.
Here, $\mathbf{W}_1$, $\mathbf{W}_2$, and $\mathbf{W}_3$ denote the weight matrices, while $\mathbf{b}_1$, $\mathbf{b}_2$, and $\mathbf{b}_3$ are the bias vectors. The operations $\text{BatchNorm}(\cdot)$ and $\text{ReLU}(\cdot)$ represent batch normalization and ReLU activation, respectively. $\text{Dropout}(\cdot, p)$ introduces regularization, and $\sigma(\cdot)$ is the sigmoid activation function.
In the NexusIndexModel II, we introduce specialized layers to enhance fake news detection. The Attention Layer selectively focuses on the most relevant embedding features, improving the model’s capacity to distinguish between real and fake news. Furthermore, the FAISSNexusIndex Layer is implemented to efficiently index and retrieve the top-$k$ nearest neighbors. This layer is seamlessly integrated into the network, enabling rapid similarity searches during training and inference. These components, along with advanced techniques like batch normalization and dropout, collectively boost model performance, as detailed in Algorithm~\ref{algo:nexusindexII}.
\subsection{Embedding Calculation}
For each article \( x \in R \cup F \), we compute the embedding \( e_x \) using a pre-trained language model \( M \). The embedding process can be described as:\begin{equation}e_x = M(x)\end{equation}
where \( M \) represents models like BERT, RoBERTa, GPT-2 \cite{radford2019language}, or DistilBERT \cite{sanh2020distilbertdistilledversionbert}. These models transform \cite{vaswani2023attentionneed} the text \( x \) into a high-dimensional vector \( e_x \in \mathbb{R}^d \), where \( d \) is the embedding dimension. Using embeddings allows the model to capture semantic similarities between articles, which is crucial for distinguishing real news from fake news. Each model contributes uniquely to capturing the semantic richness of the input text. Rather than concatenating these embeddings, our approach evaluates each model independently to determine the most effective representation for distinguishing between real and fake news. Each model has been chosen for its particular strengths—BERT and RoBERTa for deep contextual understanding, GPT-2 for capabilities and handling longer texts, and DistilBERT for efficient processing.These models represent some of the best-performing approaches in natural language processing. 

\begin{algorithm}
\caption{Proposed NexusIndexModel}
\label{algo:nexusindex}
\KwIn{Real News Dataset $R$, Fake News Dataset $F$}
\KwOut{Trained NexusIndex Model, Evaluation Metrics}
\SetKwFunction{KwInit}{Initialize}
\KwInit{Load datasets $R$ and $F$\;}
Calculate embeddings using Multi-Model: $E_R = \text{Model}(R)$, $E_F = \text{Model}(F)$\;
Index embeddings using FAISS: $I \gets \text{FAISS}(E_R)$\;
Split data into training and testing sets: $X_{\text{train}}, X_{\text{test}}, y_{\text{train}}, y_{\text{test}}$\;
Scale features: $X_{\text{train}}, X_{\text{test}} \gets \text{StandardScaler}(X_{\text{train}}, X_{\text{test}})$\;
Define neural network model: $f_\theta(x) = \sigma(W_3 \cdot \text{ReLU}(W_2 \cdot \text{ReLU}(W_1 \cdot x + b_1) + b_2) + b_3)$\;
Train model: $\theta \gets \arg\min_\theta L(f_\theta(X_{\text{train}}), y_{\text{train}})$\;
Evaluate model accuracy: $\text{Accuracy} = \frac{1}{N} \sum_{i=1}^{N} \mathbb{I}(\text{round}(f_\theta(X^{(i)}_{\text{test}})) = y^{(i)}_{\text{test}})$\;
\end{algorithm}

\subsection{Vector Database Integration}
The calculated embeddings are indexed using FAISS, which enhances the storage and retrieval of high-dimensional embeddings. This integration is crucial for efficiently managing and querying large-scale embedding data, enabling rapid similarity searches and ensuring scalability within our fake news detection framework. The FAISS index \( I \) is then constructed for the real news dataset \( R \), allowing for efficient similarity searches as described in Algorithm~\ref{algo:nexusindex}.
We create an index of embeddings for the real news dataset \( R \) using the FAISS. Let \( E_R = [e_{r_i}]^T \), where \( i \in [1, n] \), denote the matrix of embeddings for \( R \). The FAISS index \( I \) is then constructed as \( I = \text{FAISS}(E_R) \).
The index \( I \) is then constructed as \( I = \text{FAISS}(E_R) \). Indexing embeddings allows for efficient similarity searches, which is essential for real-time fake news detection. These embeddings are then indexed using, as described in Algorithm~\ref{algo:nexusindex}.
\subsection{FAISSNexusIndex Layer in Neural Network}
We propose \textbf{FAISSNexusIndex as a layer} within the neural network. This layer allows the network to efficiently retrieve and process the top-\( k \) most similar real news embeddings during training and inference.
\begin{equation} D, I = \text{FAISSNexusIndex}(e_{f_i}) \end{equation} where \( D \) are the distances and \( I \) are the indices of the top-\( k \) similar embeddings. Using FAISSNexusIndex as a layer enables the integration of efficient nearest-neighbor search directly within the neural network's forward pass, enhancing the model's ability to leverage similarity information. Integrating FAISS as both a vector database and as a neural network layer allows for efficient nearest-neighbor search directly within the model’s forward pass. This integration enhances the system’s ability to handle large-scale, high-dimensional data efficiently, ensuring rapid similarity searches and scalability, similar to the caching mechanisms discussed in \cite{monir2024efficient}.
The proposed \textbf{NexusIndexModel} algorithm~\ref{algo:nexusindexII} systematically enhances fake news detection by integrating multi-model embeddings and efficient vector database indexing. Initially, the real and fake news datasets are loaded, and embeddings are computed using various advanced language models. These embeddings capture a wide range of contextual and semantic nuances. To facilitate rapid similarity searches, the embeddings are indexed, enabling efficient retrieval of relevant articles. A neural network model is defined, comprising multiple fully connected layers with ReLU activations, and trained to minimize binary cross-entropy loss. The NexusIndexModel consists of three fully connected layers with 128, 64, and 1 units, respectively. Each layer is followed by batch normalization and dropout to prevent overfitting. ReLU activation is applied to the hidden layers, and a sigmoid activation is used for the output layer to produce binary classification probabilities. This algorithm not only optimizes the detection process but also ensures scalability and accuracy in handling large datasets.
where $\mathbf{W}_1$, $\mathbf{W}_2$, and $\mathbf{W}_3$ are weight matrices, $\mathbf{b}_1$, $\mathbf{b}_2$, and $\mathbf{b}_3$ are bias vectors, $\text{BatchNorm}(\cdot)$ represents batch normalization, $\text{ReLU}(\cdot)$ is the ReLU activation function, $\text{Dropout}(\cdot, p)$ is the dropout operation with probability $p$, and $\sigma(\cdot)$ is the sigmoid activation function. In the NexusIndexModel II, we introduce and propose several specialized layers to enhance the performance of fake news detection. Key among these innovations is the \textbf{Attention Layer}, which is applied to the embeddings before the \textbf{FAISSNexusIndex layer}. This attention mechanism selectively focuses on the most relevant features within the embeddings, thereby enhancing the model’s ability to differentiate between real and fake news articles. Subsequently, the refined embeddings are passed through the FAISSNexusIndex layer to efficiently index and retrieve the top-k nearest neighbors for each article. This indexing layer is seamlessly integrated into the model, enabling rapid similarity-based searches during both training and inference. These layers, along with fully connected neural network layers that incorporate advanced techniques such as batch normalization and dropout, collectively enhance model performance, as detailed in Algorithm algorithm~\ref{algo:nexusindexII}.
\begin{algorithm}
\caption{Proposed NexusIndexModel II for Fake News Detection}
\label{algo:nexusindexII}
\begin{algorithmic}[1]
\Require Datasets $R$, $F$, Model $M$, Neighbors $k=5$
\Ensure Trained model $\theta$, Metrics
\Statex
\State Compute embeddings: $e_x = M(x)$ for $x \in R \cup F$
\State Label embeddings: $y_r = 1$ for $r \in R$, $y_f = 0$ for $f \in F$
\State Apply attention: $a_x = \text{softmax}(W_a e_x)$
\State Index $E_R$: $I = \text{FAISS}(E_R)$
\State Retrieve top-$k$: $(D_x, I_x) = I.\text{search}(e_x, k)$
\State Define model: $\hat{y} = \sigma(W_3 \text{ReLU}(W_2 \text{ReLU}(W_1 D_x)))$
\State Minimize loss: $L(\theta) = -\frac{1}{N} \sum_{i=1}^{N} \left[y_i \log(\hat{y}_i) + (1 - y_i) \log(1 - \hat{y}_i)\right]$
\State Update $\theta$: $\theta \leftarrow \theta - \eta \nabla_\theta L(\theta)$
\State Compute metrics: Accuracy, MRR, Recall@$k$, nDCG@$k$
\State \textbf{Return} $\theta$, Metrics
\end{algorithmic}
\end{algorithm}
\subsection{Attention Mechanism and Model Training}
In \textbf{NexusIndexModel II}, an attention mechanism is proposed to prioritize the most relevant features within the embeddings. This is mathematically defined as \( a_i = \text{Attention}(s_i) \). The attention mechanism helps the model focus on significant similarities, thereby improving classification accuracy. The training process involves minimizing the binary cross-entropy loss to optimize the network parameters \( \theta \), as demonstrated in Algorithm~\ref{algo:nexusindexII}.
\subsection{Feature Scaling and Model Training}
The dataset is split into training and testing sets: \( D_{\text{train}} \) and \( D_{\text{test}} \). The features are standardized using a StandardScaler:
\begin{equation}
X_{\text{train}} = \text{StandardScaler().fit\_transform}(X_{\text{train}})
\end{equation}
\begin{equation}
X_{\text{test}} = \text{StandardScaler().transform}(X_{\text{test}})
\end{equation}
Standardizing features ensures that the neural network training process is more stable and efficient. We define a neural network \( f_\theta \) with parameters \( \theta \) to classify the news articles. The network architecture is:
\begin{equation}
f_\theta(x) = \sigma(W_3 \cdot \text{ReLU}(W_2 \cdot \text{ReLU}(W_1 \cdot x + b_1) + b_2) + b_3)
\end{equation}
where \( W_i \) and \( b_i \) are the weights and biases of the \( i \)-th layer, and \( \sigma \) is the sigmoid activation function. The training objective is to minimize the binary cross-entropy loss \( L \) over the training set:
\begin{equation}
\theta = \arg\min_\theta \sum_{i=1}^{N} L(f_\theta(x_i), y_i)
\end{equation}
where \( N \) is the number of training samples, \( x_i \) is the feature vector, and \( y_i \) is the true label. Finally, we evaluate the model accuracy on the test set:
\begin{equation}
\text{Accuracy} = \frac{1}{N} \sum_{i=1}^{N} \mathbb{I}(\text{round}(f_\theta(x_i)) = y_i)
\end{equation}
where \( \mathbb{I} \) is the indicator function.
The model's accuracy is evaluated on the test set using standard metrics, ensuring the robustness of the fake news detection system, as further elaborated in both Algorithm~\ref{algo:nexusindex} and Algorithm~\ref{algo:nexusindexII}.

Integrating FAISS as a vector database and as a neural network layer allows for efficient nearest-neighbor search directly within the model's forward pass. This integration enhances the system's ability to handle large-scale, high-dimensional data efficiently, ensuring rapid similarity searches and scalability.
\section{Evaluation}
\subsection{Datasets}
In our research, we utilize the Politifact, GossipCop, WELFake, and ABC News datasets as benchmarks to evaluate the performance of our fake news detection model \cite{shu2017fake, shu2017fakenewsnet, shu2020fake}. By benchmarking against these well-established and diverse datasets, we ensure that our model's effectiveness is thoroughly validated across multiple domains. \textbf{A Million News Headlines.} This contains data on news headlines published over the span of nineteen years, provides an overview of significant global events from 2003 to 2021 \cite{Kulkarni2018}. We \textbf{propose fine-tuning the RoBERTa model on the WELFake dataset} to enhance its ability to \textbf{label the ABCNews dataset}. After fine-tuning, the ABCNews headlines were processed using the RoBERTa tokenizer to ensure compatibility with the model's input requirements. \textbf{WELFake.} WELFake is a dataset of 72,134 news articles with 35,028 real and 37,106 fake news. For this, authors merged four popular news datasets (i.e. Kaggle, McIntire, Reuters, BuzzFeed Political) \cite{Verma2021-tr, verma2021welfake}. The details of these datasets are reported in Table \ref{table:dataset_statistics}.
\begin{figure}[h]
\small
\centering
\includegraphics[width=\linewidth]{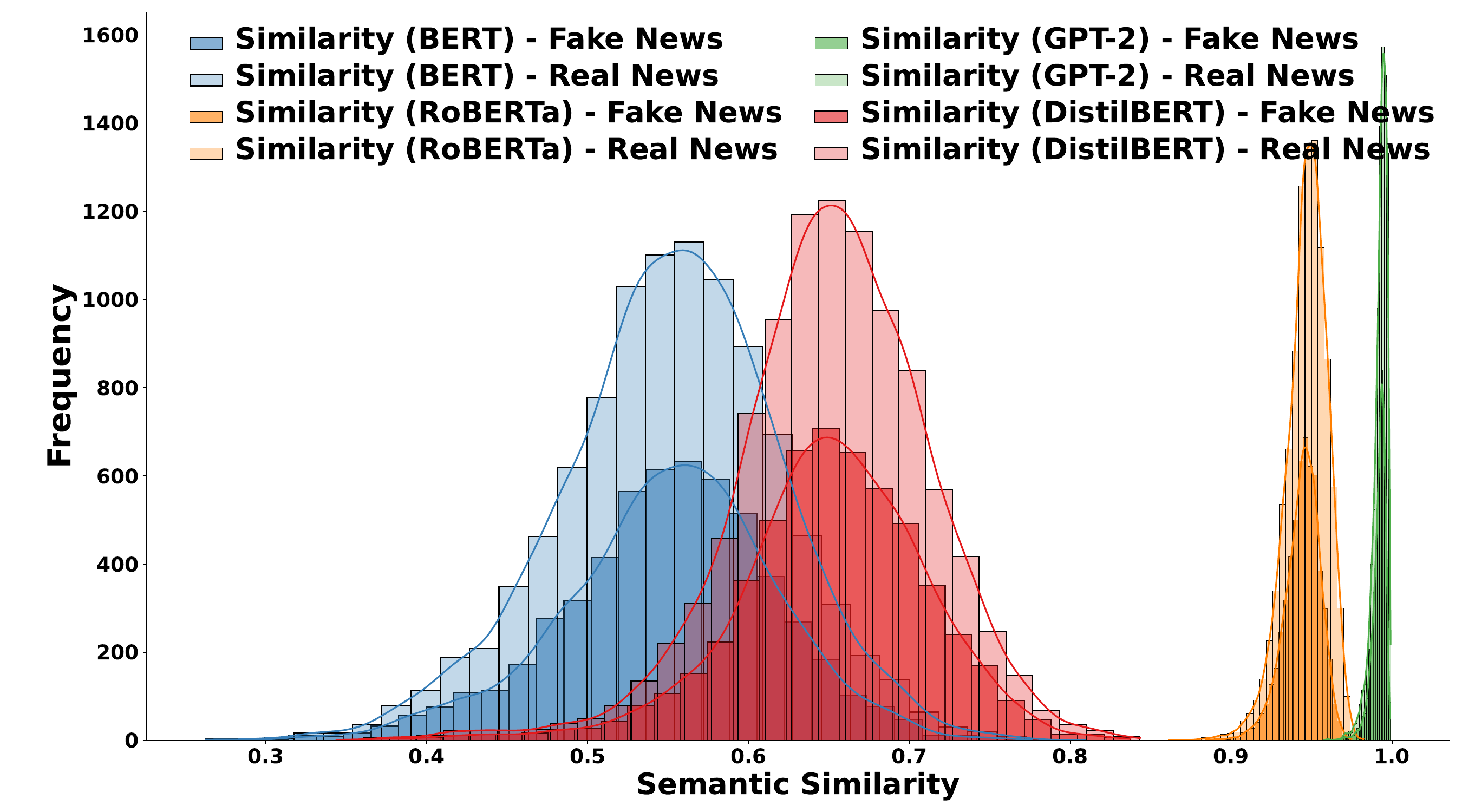}
\caption{Distribution of Semantic Similarity Scores.}
\label{fig_Semantic_Similarity}
\end{figure}
\begin{table}[h]
\small
\centering
\begin{tabular}{lcccc}
\hline
\textbf{Dataset} & \textbf{Total} & \textbf{Fake} & \textbf{Real} \\ \hline
\textbf{Labeled ABCNews (ours)}  & 1000 & 42 & 958 \\ 
WELFake & 72134 & 37106 & 35028 \\ 
A Million News & 1244184 & 0 & 1244184 \\ 
PolitiFact & 21152 & 11760 & 9392 \\ 
GossipCop & 22140 & 5323 & 16817 \\ 
\hline
\end{tabular}
\caption{Statistics for different datasets.}
\label{table:dataset_statistics} 
\end{table}
\subsection{Feature Engineering and Embedding Indexing}
\begin{figure}[h]
    \centering
    \includegraphics[width=0.40\textwidth]{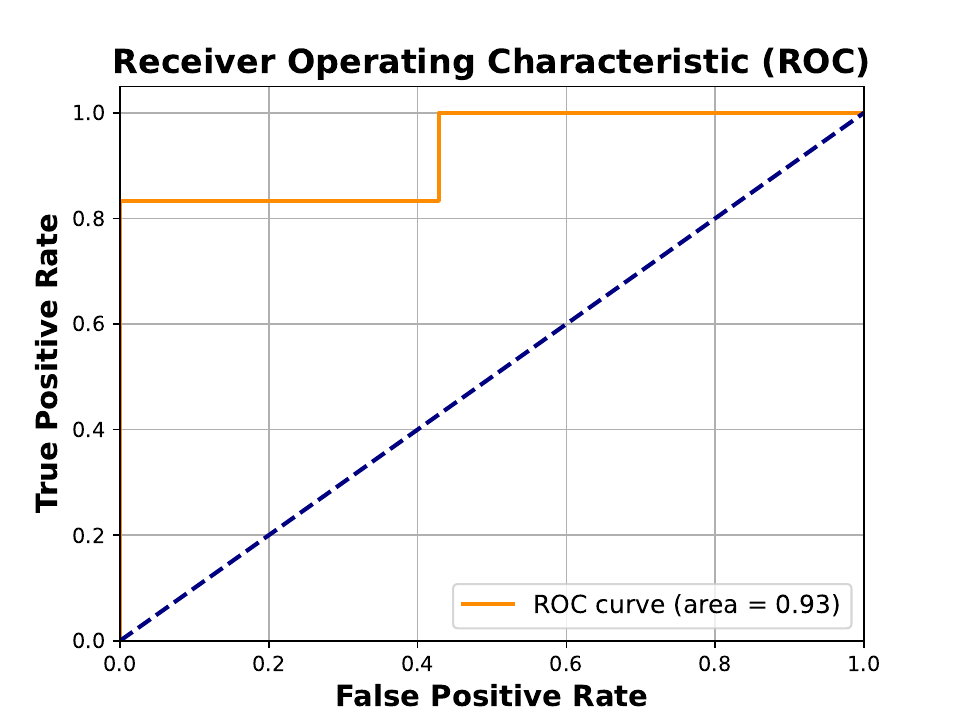}
    \caption{ROC Curve showing \textbf{NexusIndex Model's} performance in distinguishing real from fake news, with an AUC of 0.93.}
    \label{fig:fig_ROC}
\end{figure}

We utilized Multi-Model embeddings to calculate cosine similarity scores between news headlines, forming the feature set with labels for real or fake news. BERT and RoBERTa embeddings distinctly separated fake and real news, effectively capturing semantic differences. In contrast, GPT embeddings showed less variation, while DistilBERT exhibited a clear but wider distribution in similarity scores, as illustrated in   Figure~\ref{fig_Semantic_Similarity}. We utilized similarity scores from Multi-Model Embeddings as features, with news labels serving as the target variable. The data was split into training $(80\%)$ and testing $(20\%)$ sets, and standardized for consistency. To enhance fake news detection, we implemented an indexing system where real news embeddings were indexed, and the top-5 similar articles were retrieved for each fake news instance, based on L2 distance. These similarity scores, along with the embeddings, were used as input features for a neural network comprising three fully connected layers $(with 64, 32, and 1 units, respectively)$. Additionally, we integrated a novel FAISSNexusIndex layer within the neural network to optimize the retrieval process. This indexing and retrieval method aligns with the approach discussed in \cite{monir2024vectorsearch}, where semantic embeddings are used to optimize document retrieval processes. In the final experiment, we introduced an attention mechanism to focus on the most relevant features, further enhancing the model’s ability to accurately distinguish between real and fake news. The integration of vectorized embeddings for efficient processing is similar to the techniques described in \cite{monir2024veclstm}, where vectorization and database integration improve data management in neural networks.
The \textbf{NexusIndexModel II} achieved an area under the curve (AUC) of $0.89$, with a test accuracy of $85.00\%$, a precision of $88.89\%$, a recall of $80.00\%$, and an F1 score of $84.21\%$. After refining the approach, the NexusIndexModel II reached an AUC of $0.93$, with a test accuracy of $95.00\%$, a precision of $100.00\%$, a recall of 83.33\%, and an F1 score of $90.91\%$ as illustrated in Figure~\ref{fig:fig_ROC}, and further supported by the classification report for NexusIndexModel II summarized in Table~\ref{tab:classification_report}, the model demonstrates robust performance in detecting fake news. We assessed retrieval performance using MRR@10, Recall@10, and nDCG@10. RoBERTa emerged as the top performer with an nDCG@10 score of $0.0437$, indicating its relative strength in capturing semantic similarities.
\begin{table}[h]
\centering
     \begin{tabular}{lccc}
    \hline
    Class & Precision & Recall & F1-Score \\
    \hline
    Fake News & 0.93 & 1.00 & 0.97 \\
    Real News & 1.00 & 0.83 & 0.91 \\
    \hline
    Overall Accuracy & \multicolumn{3}{c}{0.95} \\
    Macro Avg & 0.97 & 0.92 & 0.94 \\
    Weighted Avg & 0.95 & 0.95 & 0.95 \\
    \hline
    \end{tabular}
    \caption{Performance Metrics for Fake News.}
    \label{tab:classification_report}
\end{table}

\begin{table}[h]
  \small
  \centering
  \begin{tabular}{lccc}
    \toprule
    \textbf{Model} & \textbf{WOS} & \textbf{NYT} & \textbf{RCV1-V2} \\
    \cmidrule(lr){2-4}
     & \textbf{Micro-F1} & \textbf{Macro-F1} & \textbf{Micro-F1} \\
    \midrule
    BERT & 85.80 & 79.20 & 78.22 \\
    BERT+HiAGM & 86.04 & 80.19 & 78.64 \\
    BERT+HTCInfoMax & 86.30 & 79.97 & 78.75 \\
    BERT+HiMatch & 86.70 & 81.06 & -- \\
    HGCLR & 87.01 & 80.30 & 78.34 \\
    HTC-CLIP & 87.86 & 81.64 & 79.22 \\
    \bottomrule
  \end{tabular}
  \caption{Performance comparison of various models on WOS, NYT, and RCV1-V2 hierarchical text classification datasets \cite{agrawal2023hierarchical, wang2022incorporating, chen2021hierarchy}.}
  \label{tab:update-model_comparison_2}
\end{table}
\begin{table*}[h!]
\centering
\begin{tabular}{lcccccc}
\hline
\textbf{Dataset} & \textbf{Model} & \textbf{Prec.} & \textbf{Rec.} & \textbf{F1} & \textbf{Acc.} & \textbf{AUC} \\ \hline
\multirow{9}{*}{Politifact} 
& SVM & 0.7460 & 0.6826 & 0.6466 & 0.6694 & 0.6826 \\  
& RFC & 0.7470 & 0.7361 & 0.7362 & 0.7406 & 0.8074 \\  
& DTC & 0.7476 & 0.7454 & 0.7450 & 0.7486 & 0.7454 \\  
& GRU-2 & 0.7083 & 0.7048 & 0.7041 & 0.7109 & 0.7896 \\  
& B-TransE & 0.7739 & 0.7658 & 0.7641 & 0.7694 & 0.7995 \\  
& KCNN & 0.7852 & 0.7824 & 0.7804 & 0.7827 & 0.8488 \\  
& KAN & 0.8687 & 0.8499 & 0.8539 & 0.8586 & 0.9197 \\  
& MPL-Full & \textbf{0.923} & 0.839 & \textbf{0.897} & 0.867 & \textbf{0.946} \\ 
& \textbf{NexusIndex (ours)} & 0.8890† & \textbf{0.8612} & 0.8724† & \textbf{0.8800} & 0.9350† \\ \hline
\multirow{9}{*}{GossipCop} 
& SVM & 0.7493 & 0.6254 & 0.5955 & 0.6643 & 0.6253 \\  
& RFC & 0.7015 & 0.6707 & 0.6691 & 0.6918 & 0.7389 \\  
& DTC & 0.6921 & 0.6922 & 0.6919 & 0.6959 & 0.6929 \\  
& GRU-2 & 0.7176 & 0.7079 & 0.7073 & 0.7180 & 0.7516 \\  
& B-TransE & 0.7369 & 0.7333 & 0.7334 & 0.7394 & 0.7995 \\  
& KCNN & 0.7483 & 0.7422 & 0.7433 & 0.7491 & 0.8125 \\  
& KAN & 0.7764 & 0.7696 & 0.7711 & 0.7766 & 0.8435 \\  
& MPL-Full & \textbf{0.869} & 0.698 & 0.565 & 0.625 & \textbf{0.921} \\ 
& \textbf{NexusIndex (ours)} & 0.7990† & \textbf{0.7805} & \textbf{0.7896} & \textbf{0.8000} & 0.8550† \\ \hline
\multirow{11}{*}{\textbf{ABC News + WELFake (ours)}} 
& LR & 0.8750 & 0.7000 & 0.7778 & 0.8000 & 0.9500 \\  
& SVM & \textbf{1.0000} & 0.7000 & 0.8235† & 0.8500† & 0.9500 \\  
& Random Forest & 0.8750 & 0.7000 & 0.7778 & 0.8000 & 0.9500 \\  
& Decision Tree & 0.5455 & 0.6000 & 0.5714 & 0.5500 & 0.9500 \\  
& GRU-2 & 0.8571 & 0.6000 & 0.7059 & 0.7500 & 0.9500 \\  
& B-TransE & 0.7778 & 0.7000 & 0.7368 & 0.7500 & 0.9500 \\  
& KCNN & 0.8333 & 0.5000 & 0.6250 & 0.7000 & 0.9500 \\  
& KAN & 0.8750 & 0.7000 & 0.7778 & 0.8000 & 0.9500 \\  
& Neural Network & 0.8750 & 0.7000 & 0.7778 & 0.8000 & 0.9500 \\  
& \textbf{NexusIndex (ours)} & \textbf{1.0000} & \textbf{0.8333} & \textbf{0.9091} & \textbf{0.9500} & \textbf{0.9600} \\ \hline
\textbf{Politifact} & \textbf{NexusIndex} & \textbf{-3.68\%} & \textbf{+2.63\%} & \textbf{-2.77\%} & \textbf{+1.50\%} & \textbf{-1.16\%} \\ 
\textbf{GossipCop} & \textbf{NexusIndex} & \textbf{-8.05\%} & \textbf{+11.81\%} & \textbf{+39.81\%} & \textbf{+17.60\%} & \textbf{-7.16\%} \\ 
\textbf{News (ours)} & \textbf{NexusIndex} & \textbf{+0\%} & \textbf{+19.04\%} & \textbf{+10.39\%} & \textbf{+11.76\%} & \textbf{+1.05\%} \\ \hline
\end{tabular}
\caption{Comparison of Baseline Models across Different Datasets. The best score is in bold, and the second-best score is marked with †.}
\label{table:baseline_models}
\end{table*}
\subsection{Baselines}
In Table~\ref{table:baseline_models}, we evaluate the performance of several baseline models on three dataset, the models tested on content-based methods that analyze the textual or visual content of news articles for fake news detection which include SVM \cite{Yang2012}, RFC \cite{kwon2013prominent}, DTC \cite{Castillo2011}, GRU-2 \cite{Ma2016}, the second category consists of knowledge-aware methods that leverage external knowledge sources, such as knowledge graphs and social context, to identify fake news. This category includes B-TransE \cite{pan2018content}, KCNN (Kernighan Convolutional Neural Network) \cite{wang2018dkn}, and KAN (Knowledge-aware Attention Network) \cite{dun2021kan}. Among these for the Politifact dataset, KAN \cite{dun2021kan} demonstrated strong performance, our proposed model, NexusIndex, surpassed these results. MPL \cite{hu2024mpl} leverages a multi-modal pre-trained model combined with a learnable prompt module. Results underscore MPL’s effectiveness in early detection scenarios. However, while MPL outperforms in terms of precision on both datasets (Politifact: 0.923 vs. 0.8890, GossipCop: 0.869 vs. 0.7990), our NexusIndex model demonstrates competitive advantages in recall and F1 score, particularly on the GossipCop dataset. For instance, NexusIndex achieves a recall of 0.8612 on Politifact, surpassing MPL’s 0.839, and a recall of 0.7805 on GossipCop, significantly higher than MPL’s 0.698. Moreover, NexusIndex maintains a balanced performance with an F1 score of 0.8724 on Politifact and 0.7896 on GossipCop, compared to MPL’s 0.897 and 0.565, respectively. While MPL excels in early detection, especially in precision, we propose NexusIndex as a more balanced solution, offering stronger recall and robustness. Its efficient similarity searches enhance its suitability for applications demanding consistent performance across diverse, large-scale datasets. In Table~\ref{tab:update-model_comparison_2}, we compare the NexusIndex model with baseline models such as BERT+HiAGM and HTC-CLIP \cite{agrawal2023hierarchical, wang2022incorporating, chen2021hierarchy}. While these models demonstrate strong performance in certain metrics like precision, NexusIndex achieves a higher recall (0.8612) on the Politifact dataset, alongside an F1 score of 0.8724.

NexusIndex's architecture, which integrates multi-model embeddings with the FAISSNexusIndex layer, facilitates more efficient similarity searches, improving classification accuracy. This design allows NexusIndex to consistently perform well across multiple datasets by better capturing semantic nuances and retrieving relevant articles. Unlike models like BERT+HiMatch and HTC-CLIP, which focus on hierarchical text classification, NexusIndex emphasizes both recall and precision, making it more adaptable and effective for large-scale fake news detection.
\begin{table}[H]
\small
\centering
\begin{tabular}{lccc}
\hline
\textbf{Metric} & \textbf{MultiModelSimNet} & \textbf{Model I} & \textbf{Model II} \\
\hline
\textbf{Accuracy} & 0.7500 & 0.8500 & \textbf{0.9500} \\
\textbf{Precision} & 0.7500 & 0.8000 & 1.0000 \\
\textbf{Recall} & 0.6667 & 0.8889 & 0.8333 \\
\textbf{F1 Score} & 0.7059 & 0.8421 & 0.9091 \\
\textbf{R@100} & N/A & N/A & 1.0  \\
\textbf{nDCG@10} & N/A & N/A & 0.207 \\
\hline
\end{tabular}
\caption{Comparison of proposed NexusIndexModel Performance.}
\label{tab:combined_comparison}
\end{table}
\subsection{Ablation Study}
We compare the performance of our two proposed models: \textbf{NexusIndexModel I} and \textbf{NexusIndexModel II}. The comparison focuses on key metrics such as Accuracy, Recall at 100 (R@100), and Normalized Discounted Cumulative Gain at 10 (nDCG@10), as summarized in Table \ref{tab:combined_comparison}.
\textbf{Accuracy:} Both models demonstrate high accuracy, correctly classifying 95\% of the data in both training and testing phases. This indicates that the core architecture is robust across different datasets. \textbf{R@100:} The NexusIndex Models II achieves perfect recall at 100, effectively retrieving all relevant documents within the top 100 ranks. This improvement highlights the model's enhanced ability to prioritize relevant data in a larger pool, which is crucial for applications requiring high recall.
\textbf{nDCG@10:} The nDCG@10 score of 0.207 for the NexusIndex Models II indicates its effectiveness in ranking relevant documents within the top 10. While there is room for improvement, this metric shows that the model can reasonably rank relevant items close to the top. By focusing on classification accuracy and retrieval effectiveness, this ablation study highlights the impact of indexing, dropout layers, and attention mechanisms on performance metrics. Initially, similarity scores were calculated using Multi-Model embeddings and served as input features for a neural network. In the second experiment, we introduced an improved neural network with indexing and dropout layers, significantly boosting recall and F1 score. Finally, the novel \textbf{NexusIndexModel II} was proposed, integrating an attention mechanism to prioritize crucial features and a \textbf{FAISSNexusIndex layer} for efficient nearest-neighbor searches, leading to superior performance as shown in Table ~\ref{tab:combined_comparison}.
\section{Conclusion and Future Works}
In this article, we proposed and evaluated two innovative models—NexusIndexModel and NexusIndexModel II—for enhancing fake news detection. Both models leverage state-of-the-art NLP techniques, including Multi-Model embeddings, and a novel FAISSNexusIndex layer for efficient and precise similarity searches directly within the neural network architecture. NexusIndexModel II demonstrated a classification accuracy of $95\%$ and achieved perfect recall, highlighting its robustness in both classification and retrieval tasks. The integration of advanced vector database techniques and attention mechanisms significantly contributed to these results, emphasizing the models' ability to handle large-scale, high-dimensional data efficiently.
\bibliography{main-2025}

\begin{thebibliography}{36}
\providecommand{\natexlab}[1]{#1}

\bibitem[{Agrawal et~al.(2023)Agrawal, Kumar, Bhatt, and Agarwal}]{agrawal2023hierarchical}
Agrawal, N.; Kumar, S.; Bhatt, P.; and Agarwal, T. 2023.
\newblock Hierarchical Text Classification Using Contrastive Learning Informed Path Guided Hierarchy.
\newblock In \emph{ECAI 2023}, 19--26. IOS Press.

\bibitem[{Bekker and Davis(2020)}]{bekker2020learning}
Bekker, J.; and Davis, J. 2020.
\newblock Learning from positive and unlabeled data: A survey.
\newblock \emph{Machine Learning}, 109(4): 719--760.

\bibitem[{Castillo, Mendoza, and Poblete(2011)}]{Castillo2011}
Castillo, C.; Mendoza, M.; and Poblete, B. 2011.
\newblock Information credibility on twitter.
\newblock In \emph{Proceedings of the 20th International Conference on World Wide Web}, 675--684. ACM.

\bibitem[{Chen et~al.(2021)Chen, Ma, Lin, and Yan}]{chen2021hierarchy}
Chen, H.; Ma, Q.; Lin, Z.; and Yan, J. 2021.
\newblock Hierarchy-aware label semantics matching network for hierarchical text classification.
\newblock In \emph{Proceedings of the 59th Annual Meeting of the Association for Computational Linguistics and the 11th International Joint Conference on Natural Language Processing (Volume 1: Long Papers)}, 4370--4379.

\bibitem[{Devlin et~al.(2018)Devlin, Chang, Lee, and Toutanova}]{devlin2018bert}
Devlin, J.; Chang, M.-W.; Lee, K.; and Toutanova, K. 2018.
\newblock Bert: Pre-training of deep bidirectional transformers for language understanding.
\newblock \emph{arXiv preprint arXiv:1810.04805}.

\bibitem[{Douze et~al.(2024)Douze, Guzhva, Deng, Johnson, Szilvasy, Mazaré, Lomeli, Hosseini, and Jégou}]{douze2024faiss}
Douze, M.; Guzhva, A.; Deng, C.; Johnson, J.; Szilvasy, G.; Mazaré, P.-E.; Lomeli, M.; Hosseini, L.; and Jégou, H. 2024.
\newblock The Faiss library.

\bibitem[{Dun et~al.(2021)Dun, Tu, Chen, Hou, and Yuan}]{dun2021kan}
Dun, Y.; Tu, K.; Chen, C.; Hou, C.; and Yuan, X. 2021.
\newblock Kan: Knowledge-aware attention network for fake news detection.
\newblock In \emph{Proceedings of the AAAI conference on artificial intelligence}, volume~35, 81--89.

\bibitem[{Gao and Callan(2022)}]{gao2022long}
Gao, L.; and Callan, J. 2022.
\newblock Long Document Re-ranking with Modular Re-ranker.
\newblock In \emph{Proceedings of the 45th International ACM SIGIR Conference on Research and Development in Information Retrieval}, 2371--2376.

\bibitem[{Hu et~al.(2024)Hu, Wang, Jia, Liao, and Zhou}]{hu2024mpl}
Hu, W.; Wang, Y.; Jia, Y.; Liao, Q.; and Zhou, B. 2024.
\newblock A Multi-modal Prompt Learning Framework for Early Detection of Fake News.
\newblock In \emph{Proceedings of the International AAAI Conference on Web and Social Media}, volume~18, 651--662.

\bibitem[{Johnson, Douze, and J{\'e}gou(2019)}]{Johnson2019}
Johnson, J.; Douze, M.; and J{\'e}gou, H. 2019.
\newblock Billion-scale similarity search with GPUs.
\newblock In \emph{IEEE Transactions on Big Data}.

\bibitem[{Kulkarni(2018)}]{Kulkarni2018}
Kulkarni, R. 2018.
\newblock A Million News Headlines.
\newblock Datestamped Article Headlines.

\bibitem[{Kwon et~al.(2013)Kwon, Cha, Jung, Chen, and Wang}]{kwon2013prominent}
Kwon, S.; Cha, M.; Jung, K.; Chen, W.; and Wang, Y. 2013.
\newblock Prominent features of rumor propagation in online social media.
\newblock In \emph{2013 IEEE 13th international conference on data mining}, 1103--1108. IEEE.

\bibitem[{Liu et~al.(2019)Liu, Ott, Goyal, Du, Joshi, Chen, Levy, Lewis, Zettlemoyer, and Stoyanov}]{liu2019roberta}
Liu, Y.; Ott, M.; Goyal, N.; Du, J.; Joshi, M.; Chen, D.; Levy, O.; Lewis, M.; Zettlemoyer, L.; and Stoyanov, V. 2019.
\newblock Roberta: A robustly optimized bert pretraining approach.
\newblock \emph{arXiv preprint arXiv:1907.11692}.

\bibitem[{Ma et~al.(2016)Ma, Gao, Mitra, Kwon, Jansen, Wong, and Cha}]{Ma2016}
Ma, J.; Gao, W.; Mitra, P.; Kwon, S.; Jansen, B.~J.; Wong, K.-F.; and Cha, M. 2016.
\newblock Detecting Rumors from Microblogs with Recurrent Neural Networks.
\newblock In \emph{Proceedings of the Twenty-Fifth International Joint Conference on Artificial Intelligence (IJCAI)}, 3818--3824. AAAI.

\bibitem[{Monir et~al.(2024)Monir, Lau, Yang, and Zhao}]{monir2024vectorsearch}
Monir, S.~S.; Lau, I.; Yang, S.; and Zhao, D. 2024.
\newblock VectorSearch: Enhancing Document Retrieval with Semantic Embeddings and Optimized Search.
\newblock \emph{arXiv preprint arXiv:2409.17383}.

\bibitem[{Monir and Zhao(2024{\natexlab{a}})}]{monir2024efficient}
Monir, S.~S.; and Zhao, D. 2024{\natexlab{a}}.
\newblock Efficient Feature Extraction for Image Analysis through Adaptive Caching in Vector Databases.
\newblock In \emph{2024 7th International Conference on Information and Computer Technologies (ICICT)}, 193--198. IEEE.

\bibitem[{Monir and Zhao(2024{\natexlab{b}})}]{monir2024veclstm}
Monir, S.~S.; and Zhao, D. 2024{\natexlab{b}}.
\newblock VecLSTM: Trajectory Data Processing and Management for Activity Recognition through LSTM Vectorization and Database Integration.
\newblock \emph{arXiv preprint arXiv:2409.19258}.

\bibitem[{Pan et~al.(2018)Pan, Pavlova, Li, Li, Li, and Liu}]{pan2018content}
Pan, J.~Z.; Pavlova, S.; Li, C.; Li, N.; Li, Y.; and Liu, J. 2018.
\newblock Content based fake news detection using knowledge graphs.
\newblock In \emph{The Semantic Web--ISWC 2018: 17th International Semantic Web Conference, Monterey, CA, USA, October 8--12, 2018, Proceedings, Part I 17}, 669--683. Springer.

\bibitem[{Radford et~al.(2019)Radford, Wu, Child, Luan, Amodei, Sutskever et~al.}]{radford2019language}
Radford, A.; Wu, J.; Child, R.; Luan, D.; Amodei, D.; Sutskever, I.; et~al. 2019.
\newblock Language models are unsupervised multitask learners.
\newblock \emph{OpenAI blog}, 1(8): 9.

\bibitem[{Rahimi et~al.(2021)Rahimi, Fouladi-Fard, Aali, Shahryari, Rezaali, Ghafouri, Ghalhari, Asadi-Ghalhari, Farzinnia, Gea et~al.}]{rahimi2021bidirectional}
Rahimi, N.~R.; Fouladi-Fard, R.; Aali, R.; Shahryari, A.; Rezaali, M.; Ghafouri, Y.; Ghalhari, M.~R.; Asadi-Ghalhari, M.; Farzinnia, B.; Gea, O.~C.; et~al. 2021.
\newblock Bidirectional association between COVID-19 and the environment: a systematic review.
\newblock \emph{Environmental Research}, 194: 110692.

\bibitem[{Sanh et~al.(2020)Sanh, Debut, Chaumond, and Wolf}]{sanh2020distilbertdistilledversionbert}
Sanh, V.; Debut, L.; Chaumond, J.; and Wolf, T. 2020.
\newblock DistilBERT, a distilled version of BERT: smaller, faster, cheaper and lighter.
\newblock arXiv:1910.01108.

\bibitem[{Shu et~al.(2017{\natexlab{a}})Shu, Sliva, Wang, Tang, and Liu}]{Shu2017}
Shu, K.; Sliva, A.; Wang, S.; Tang, J.; and Liu, H. 2017{\natexlab{a}}.
\newblock Fake News Detection on Social Media: A Data Mining Perspective.
\newblock \emph{ACM SIGKDD Explorations Newsletter}, 19(1): 22--36.

\bibitem[{Shu et~al.(2017{\natexlab{b}})Shu, Sliva, Wang, Tang, and Liu}]{shu2017fake}
Shu, K.; Sliva, A.; Wang, S.; Tang, J.; and Liu, H. 2017{\natexlab{b}}.
\newblock Fake News Detection on Social Media: A Data Mining Perspective.
\newblock In \emph{Proceedings of the 2017 ACM SIGKDD international conference on knowledge discovery and data mining}, 22--25. ACM.

\bibitem[{Shu et~al.(2020)Shu, Wang, Lee, and Liu}]{shu2020fake}
Shu, K.; Wang, S.; Lee, D.; and Liu, H. 2020.
\newblock Fake news detection on social media: A data mining perspective.
\newblock \emph{ACM SIGKDD Explorations Newsletter}, 19(1): 22--36.

\bibitem[{Shu, Wang, and Liu(2017)}]{shu2017fakenewsnet}
Shu, K.; Wang, S.; and Liu, H. 2017.
\newblock FakeNewsNet: A data repository with news content, social context and dynamic information for studying fake news on social media.
\newblock In \emph{Companion Proceedings of the 2017 ACM Conference on Computer Supported Cooperative Work and Social Computing (CSCW)}, 243--246. ACM.

\bibitem[{Vaswani et~al.(2023)Vaswani, Shazeer, Parmar, Uszkoreit, Jones, Gomez, Kaiser, and Polosukhin}]{vaswani2023attentionneed}
Vaswani, A.; Shazeer, N.; Parmar, N.; Uszkoreit, J.; Jones, L.; Gomez, A.~N.; Kaiser, L.; and Polosukhin, I. 2023.
\newblock Attention Is All You Need.
\newblock arXiv:1706.03762.

\bibitem[{Verma et~al.(2021)Verma, Agrawal, Amorim, and Prodan}]{verma2021welfake}
Verma, P.~K.; Agrawal, P.; Amorim, I.; and Prodan, R. 2021.
\newblock WELFake: word embedding over linguistic features for fake news detection.
\newblock \emph{IEEE Transactions on Computational Social Systems}, 8(4): 881--893.

\bibitem[{Verma, Agrawal, and Prodan(2021)}]{Verma2021-tr}
Verma, P.~K.; Agrawal, P.; and Prodan, R. 2021.
\newblock {WELFake} dataset for fake news detection in text data.

\bibitem[{Wang et~al.(2018)Wang, Zhang, Xie, and Guo}]{wang2018dkn}
Wang, H.; Zhang, F.; Xie, X.; and Guo, M. 2018.
\newblock DKN: Deep knowledge-aware network for news recommendation.
\newblock In \emph{Proceedings of the 2018 world wide web conference}, 1835--1844.

\bibitem[{Wang et~al.(2022{\natexlab{a}})Wang, Yang, Huang, Jiao, Yang, Jiang, Majumder, and Wei}]{wang2022simlm}
Wang, L.; Yang, N.; Huang, X.; Jiao, B.; Yang, L.; Jiang, D.; Majumder, R.; and Wei, F. 2022{\natexlab{a}}.
\newblock Simlm: Pre-training with representation bottleneck for dense passage retrieval.
\newblock \emph{arXiv preprint arXiv:2207.02578}.

\bibitem[{Wang et~al.(2022{\natexlab{b}})Wang, Wang, Huang, Sun, and Wang}]{wang2022incorporating}
Wang, Z.; Wang, P.; Huang, L.; Sun, X.; and Wang, H. 2022{\natexlab{b}}.
\newblock Incorporating hierarchy into text encoder: a contrastive learning approach for hierarchical text classification.
\newblock \emph{arXiv preprint arXiv:2203.03825}.

\bibitem[{Whitehouse et~al.(2022)Whitehouse, Weyde, Madhyastha, and Komninos}]{Whitehouse2022}
Whitehouse, C.; Weyde, T.; Madhyastha, P.; and Komninos, N. 2022.
\newblock Evaluation of Fake News Detection with Knowledge-Enhanced Language Models.
\newblock In \emph{Proceedings of the Sixteenth International AAAI Conference on Web and Social Media}. AAAI.

\bibitem[{Yang and Liu(2012)}]{Yang2012}
Yang, Y.; and Liu, X. 2012.
\newblock Support Vector Machines for Large-Scale Data: Applications in Text Categorization and Handwriting Recognition.
\newblock \emph{Proceedings of the IEEE}, 98(8): 1348--1358.

\bibitem[{Zhang et~al.(2022)Zhang, Wu, Zhang, Zhu, Lin, Zhang, Sun, He, Mueller, Manmatha et~al.}]{zhang2022resnest}
Zhang, H.; Wu, C.; Zhang, Z.; Zhu, Y.; Lin, H.; Zhang, Z.; Sun, Y.; He, T.; Mueller, J.; Manmatha, R.; et~al. 2022.
\newblock Resnest: Split-attention networks.
\newblock In \emph{Proceedings of the IEEE/CVF conference on computer vision and pattern recognition}, 2736--2746.

\bibitem[{Zhang et~al.(2024)Zhang, Zhang, Zhou, Huang, and Li}]{zhang2024reinforced}
Zhang, L.; Zhang, X.; Zhou, Z.; Huang, F.; and Li, C. 2024.
\newblock Reinforced adaptive knowledge learning for multimodal fake news detection.
\newblock In \emph{Proceedings of the AAAI Conference on Artificial Intelligence}, volume~38, 16777--16785.

\bibitem[{Zhou et~al.(2024)Zhou, Shen, Geng, Tao, Shen, Long, Xu, and Jiang}]{zhou2024fine}
Zhou, Y.; Shen, T.; Geng, X.; Tao, C.; Shen, J.; Long, G.; Xu, C.; and Jiang, D. 2024.
\newblock Fine-grained distillation for long document retrieval.
\newblock In \emph{Proceedings of the AAAI Conference on Artificial Intelligence}, volume~38, 19732--19740.

\end{thebibliography}

\end{document}